# Magnetospheres of Terrestrial Exoplanets and Exomoons: Implications for Habitability and Detection

James Green[1], Scott Boardsen[2], and Chuanfei Dong[3]

[1]NASA Headquarters.

[2]University of Maryland Baltimore County.

[3]Department of Astrophysical Sciences and Princeton Plasma Physics Laboratory, Princeton University.

**Abstract**
Characterizing habitable exoplanets and/or their moons is of paramount importance. Here we show the results of our magnetic field topological modeling which demonstrate that terrestrial exoplanet-exomoon coupled magnetospheres work together to protect the early atmospheres of both the exoplanet and the exomoon. When exomoon magnetospheres are within the exoplanet's magnetospheric cavity, the exomoon magnetosphere acts like a protective magnetic bubble providing an additional magnetopause confronting the stellar winds when the moon is on the dayside. In addition, magnetic reconnection would create a critical pathway for the atmosphere exchange between the early exoplanet and exomoon. When the exomoon's magnetosphere is outside of the exoplanet's magnetosphere it then becomes the first line of defense against strong stellar winds, reducing exoplanet's atmospheric loss to space. A brief discussion is given on how this type of exomoon would modify radio emissions from magnetized exoplanets.

## 1. Introduction

Habitable terrestrial planets are those occupying orbits around a star that can maintain surface liquid water under a supporting atmosphere [Kasting *et al.*, 1993]. The study of the evolution of planetary atmospheres has been an important research topic for the last several years in an effort to determine the most important factors in creating a habitable environment for an exoplanet. A planet found in the habitable zone (HZ) around a star does not necessarily mean that the planet is habitable since stellar activity cannot be neglected [e.g., Dong *et al.*, 2017, 2018a]. For example, it is well known that stellar emissions in the X-ray and extreme ultraviolet (EUV) leads to enhanced ionization and inflation of planetary ionospheres and atmospheres leading to atmospheric loss [e.g., Airapetian *et al.*, 2017]. Magnetically active stars produce very intense stellar flares that are often, but not always, accompanied by a coronal mass ejection (CME) and stellar winds leading to more atmospheric loss as observed at Mars [e.g., Jakosky *et al.*, 2015; Luhmann *et al.*, 2017]. When the Martian dynamo shut down ~4.1 Ga and Mars lost its global magnetosphere [Lillis *et al.*, 2013], the intense solar wind and radiation ravished its atmosphere resulting in ocean evaporation and atmospheric loss transforming Mars from an early warmer and wetter world to a cold and dry planet with an average surface pressure of only 6 mbar [Jakosky *et al.*, 2018; Dong *et al.*, 2018b]. Simply stated, preservation of an atmosphere is one of the chief ingredients for surface habitability.

It has been shown that young stars, particularly one solar mass and smaller, produce extremes such as stellar flares and CMEs that lead to planetary atmospheric loss and that these extremes are closely related to the stellar rotation rate in addition to mass. Johnstone *et al.* [2015] found that almost all solar mass stars have converged to slow rotators by 500 Myr after formation producing slower stellar wind. Compared with solar-type stars, it takes a longer time for M-type stars (with lower masses) to slow down. M-type stars also keep magnetically active for a longer period of time. Therefore, the ravaging of planetary atmospheres in the young solar system due to extreme solar radiation and particle fluxes is believed to be a significant factor for our understanding of how an exoplanet will develop and maintain an atmosphere, which is a critical element of a habitable environment. Meanwhile, recent studies show that planetary magnetic fields may protect planets from atmospheric losses [Owen and Adams, 2014; Gallet *et al.*, 2017; Egan *et al.*, 2019; Dong *et al.*, 2020], indicating planetary magnetic fields play an important role in planetary habitability.

Another factor that should be considered with respect to the habitability of a terrestrial exoplanet, unrecognized until this paper, concerns the magnetic characteristics of an associated exomoon. With the existence of exoplanets well established, one of the next frontiers is the discovery of exomoons. Today there are a number of exomoon candidates waiting to be confirmed [Teachey and Kipping 2018; Moraes and Vieira Neto, 2020]. Since it is without a doubt that they must exist around some exoplanets, it is important to examine what role, if any, they would have in creating an environment that contributes to the habitability of their host planet.

Although speculated for several decades, only recently have scientists determined that our Moon had an extensive magnetosphere for several hundred million years soon after it was formed [Mighani *et al.*, 2020]. Recently, Green *et al.* [2020] investigated the expected magnetic topology of the early Earth-Moon magnetospheres and found that they would couple in such a way as to protect the atmosphere of both the Earth and Moon. Assuming similar formation processes for terrestrial planets and their moons, how would these two magnetospheres interact, and what protection would such a combined magnetosphere afford to the atmospheres of early exoplanets and their moons orbiting young stars is the subject of this paper.

**2. Planet-Moon Formation**
The formation of exomoons should naturally be based on an understanding of how moons in our own solar system must have formed. The prevailing model for the generation of our Moon is the giant impact hypothesis in which a differentiated Mars sized protoplanet impacted the proto-Earth during the early formation of the solar system [Canup and Asphaug, 2001]. It is expected that extensive damage was done to the proto-Earth even to the point of eliminating its primordial atmosphere [Cameron and Benz, 1991]. The resulting debris formed the Earth and the Moon, ending with the Moon lying just outside the Roche limit of ~2.9 Earth Radii ($R_E$). Soon after formation, internal dynamos were being generated in both the Earth and the Moon [Green *et al.*, 2020]. Once formed, the Moon slowly moved away from the Earth due to the dissipation of tidal energy to its present position of 60 $R_E$.

It is believed that early magnetospheric protection is critical for the outgassing Earth to build back its secondary atmosphere and maintain it throughout periods of intense solar radiation and particle fluxes emanating from our young Sun. Intrinsic magnetic fields are generated by large dynamos carrying electric currents via convection inside planetary bodies [Stevenson, 2003]. Molten cores within spinning planets are subject to Coriolis forces and are an important element in maintaining core convection. The physical system behind the source electric currents generated inside our Moon producing its magnetosphere are still under debate [Tikoo *et al.*, 2017; Mighani *et al.*, 2020] but the Apollo lunar surface samples show its magnitude and duration. Magnetospheres in the solar system can be quite complex with multiple magnetic moments such as Uranus [Jarmak *et al.*, 2020]. It is not known what the lunar magnetosphere looked like but even a simple electric current system generates a dipole-like field.

In the above scenario of planet-moon formation it is expected that both objects upon initial accretion would be fast rotators and tidal forces would come into play while each body is still undergoing differentiation. It is only over time that the planet's moon becomes tidally locked. For those planetary bodies that are able to maintain a dynamo they will generate a magnetosphere.

### 3. Magnetospheres

Planets are subjected to a continuous plasma outflow from the host stars. The boundary region between this stellar outflow or wind and a planetary magnetosphere contains an electrical current layer known as the magnetopause. The dynamic pressure of the stellar wind is balanced by the static pressure of the planet's magnetic field at the magnetopause. For strong magnetospheres, this standoff distance to the magnetopause must be several planetary radii. Weak intrinsic magnetic fields, those in which the resulting magnetopause would be located below an induced magnetosphere boundary, would actually enhance atmospheric escape [Egan *et al.*, 2019]. These conditions typically occur when the magnetopause is within one radius of the planetary body in a stellar wind. In this study we consider only strongly magnetized exoplanets producing significant shielding similar in nature to the Earth and our own Moon as in Green *et al.* [2020]. The location of the Earth's magnetopause is typically at 10 $R_E$, which has been compressed by the solar wind such that the total magnetic field strength (**B**) at the subsolar point is about 2.5 times as large as the field would have been without the solar wind. This increase in the magnetic field at the magnetopause is due to a current that arises to cancel the planet's magnetic field outside the magnetopause. The result is that the magnetopause produces a magnetic cavity that can protect the planet from direct stellar wind interactions.

The plasma outflow from a host star typically contains a stellar magnetic field. Under the conditions where the direction of the magnetic field in the stellar wind and the direction of the planet's magnetic field at the magnetopause are opposite, a dynamic process called magnetic reconnection occurs in which the location and topology of the magnetopause changes quickly by erosion of the magnetopause location moving the magnetopause inward until a new equilibrium is established. For this first order analysis, the typical interior magnetospheric currents, such as the ring current and Chapman currents, will be ignored and we will concentrate on stellar winds interacting with the exoplanet-exomoon coupled magnetospheres working together as a single but complex structure contained by a magnetopause providing protection from atmospheric erosion.

### 4. Modeling Results

Using the technique developed by Green *et al.* [2020], we present simplified dipole magnetic field topological modeling confined within a paraboloidal shaped magnetopause to show, qualitatively, how the exoplanet-exomoon magnetospheres are coupled together as the exomoon moves outward over time in a stellar wind environment of a host star. As illustrative of the resulting topology (Figure 1), the early dipole moments of the Earth-Moon system will be used assuming that the internal dynamos become operational soon after the formation of the exoplanet and exomoon. The magnetic fields of the exoplanet and the exomoon are approximated by dipoles with the Earth-Moon magnetic moments directed in the normal (Z) direction of the ecliptic plane of $-3\times10^4$ nT $R_p^3$ and $\pm1\times10^3$ nT $R_p^3$ respectively (where $R_p$ is the exoplanet radii). This is a reasonable first order approach since, in our solar system, many planets generate magnetospheres that are dipolar in nature due to the fact that a current inside a planet will typically produce the most power into its fundamental frequency (manifesting into a magnetic dipole configuration) and much less power in the higher harmonics. The Earth's magnetic field roughly aligns with its spin axis but also periodically flips orientation. Therefore, the modeled magnetic topology of the exoplanet and exomoon will take into account the expected extremes of both, dipole aligned and antialigned configurations.

Figure 1 shows the results of the evolution of the magnetic topology of an exoplanet-exomoon over a period of time delineated by the exomoon moving away from the exoplanet which is expected to occur over 100's of millions of years. The corresponding top and bottom panels have the exomoon at the same locations (4, 8 and 18 $R_p$), respectively, and orientation (exomoon at the sub-stellar point) and under the two conditions of their dipoles aligned (top panels) and dipoles antialigned (bottom panels). The horizontal axis establishes the star-planet/moon direction with the vertical axis lying perpendicular to the orbital plane. The starting location of the exomoon at 4 $R_p$ is taken to be outside the Roche limit. In all these configurations, the coupled magnetospheres provide a significant buffer from intense stellar winds, reducing atmospheric loss to space and improving the chances of creating a habitable exoplanet. This scenario may be significant when considering terrestrial exoplanets with exomoons orbiting M-type stars given the extreme stellar environment in their close-in HZs.

The case of dipoles aligned topology (top panels of Figure 1) depicts how the exomoon's magnetosphere coupling evolves over time. At 4 $R_p$ and 8 $R_p$ configurations the exomoon magnetosphere lies within the exoplanet magnetosphere. When comparing the exomoon located at 8 $R_p$ and 18 $R_p$, the key difference between these two panels illustrates how the exomoon emerges from the exoplanet's magnetosphere with the upper right panel showing no exoplanet field lines configure to drape over the exomoon dayside. In contrast, for dipoles antialigned (bottom panels of Figure 1) show that the reconnected field lines in the polar regions of the exomoon and exoplanet are still maintained with strong coupling between these magnetospheres. Significant reconnection occurs producing an environment in which the exomoon and exoplanet have many pole-to-pole magnetic field lines allowing exchange of atmospheric material originating in their ionospheres and atmospheres. It is also important to note when the dipoles are aligned the magnetopause is extended by about 2 $R_p$ compared with the antialigned cases. Figure 1 illustrates an extremely important feature of the exomoon magnetosphere and its relationship to the exoplanet magnetopause. It is clear that an extended magnetopause arises forcing stellar wind interactions to take into account the exomoon's magnetic field.

To better quantify this interaction, Figure 2 shows the intensity of the coupled magnetic field from the exoplanet's surface, near the pole, to beyond the magnetopause along the planet-star line. Panels A and B are with the exomoon at 4 $R_p$ and panels C and D are at 12 $R_p$ for the cases of dipole aligned (A and C) and antialigned (B and D). The 12 $R_p$ location for the exomoon was selected to be just outside the exoplanet's magnetosphere when the exomoon has no magnetic field showing the maximum effect of the combined fields. The rapid enhancement (shown in red) at the end of the traces for all panels of Figure 2 indicates the model calculated magnetopause locations with the expected magnetic field contribution from a magnetopause current. As illustrated in Figure 2, the magnetospheres of the exoplanet and exomoon, in these configurations do provide a substantial barrier to the stellar wind. The extended magnetopause configuration in panel A compared to panel B is a result of the exoplanet's magnetosphere rapping around the exomoon. No difference in the magnetopause location is seen in panel C or D, since it is due solely to the exomoon, but a clear magnetic intensity drop occurs in the antialigned case (panel D) due to reconnection of field lines between the magnetospheres.

In order to qualitatively investigate other local time configurations under varying stellar wind conditions, Figure 3 provides an orbit plane projection (X-Y plane) of the exomoon (red curve) orbiting at 18 $R_p$ and exoplanet (blue curve) magnetopauses which are calculated independent of each other (coupling not modeled). Figure 3A is for a stellar wind of 2 nPa (an average solar wind pressure at 1 AU) and in panel B of 20 nPa (extreme conditions) an order of magnitude above the average. Plotting the resulting magnetopauses as if uncoupled (alone in the stellar wind) clearly shows the locations of the coupling zones in each of the configurations. Higher stellar wind pressure produces magnetospheres reduced in size. However, even with the exomoon orbiting at 18 $R_p$ the magnetospheres must still couple in significant ways in the regions where the blue and red curves overlap. A full magnetohydrodynamic model would be necessary to fully examine the details of how these magnetospheres would couple which is beyond the scope of this analysis.

**5. Discussion**
Characterizing Earth-sized habitable exomoons has already been considered [Kaltenegger, 2010]. It is expected that in extrasolar systems with Jovian-type planets their moons would probably be embedded within the giant planet's magnetosphere and therefore protected from stellar wind extremes, but what about terrestrial planets which form moons in the terrestrial domain much closer to the host star?

The search for exomoons is just beginning with a number of techniques being implemented [Kipping, 2011b]. The discovery of exomoons around terrestrial exoplanets will help determine if they are rare or common providing an understanding of how special our own Moon may be. Ward and Brownlee [2000] and Benn [2001] proposed that our Moon plays a number of critical roles in keeping the Earth "uncommonly" habitable since it causes lunar tides, stabilizes the Earth's spin axis, and slows the Earth's rotation rate, all believed to be important for the development of complex life. As proposed in Green *et al.* [2020] our Moon may have also provided a magnetic shield at a time when the early Earth was developing an atmosphere through outgassing and exposed to extremes in space weather produced by our young Sun. In a similar manner, terrestrial exoplanets with a significant exomoon may also have these same advantages and may harbor complex life thereby focusing our future research attention to these special objects is of greater importance.

In addition to investigating techniques and developing the capability to directly find an exomoon [e.g., Kipping, 2011a; Heller *et al.*, 2014], there has been a number of efforts to try and find exoplanets with magnetospheres. Based on the solar system, one of the most well-known wave phenomena for planets with magnetospheres is the release of escaping radio emissions generated by the cyclotron maser instability (CMI) mechanism that derives their named based on the frequency range of the observed emission. It is well known that this type of emission is closely related to the local gyrofrequency above a planet's aurora and therefore provides important clues to the presence of a planet's magnetic field and its strength. For example, the Earth's intense auroral-related radio emission is called Auroral Kilometric Radiation (AKR) and the Jovian Decametric (DAM) emissions are CMI related emissions from Jupiter's auroral zone.

It is believed that the intense auroral related radio emission is the best indicator of planetary magnetospheres [e.g., Bastian *et al.*, 2000; Zarka *et al.*, 2001]. The CMI generated radio emissions produce intense radiation perpendicular to the local magnetic field but the resulting emission cone can be filled-in by refraction or hollow. For instance, the emission cone of AKR has been observed to be relatively well filled in [Green and Gallagher, 1985] at higher frequencies and may be hollow at the lower frequencies [Calvert, 1986] while the Jovian DAM emissions produce hollow emission cones [Green, 1984] as illustrated in Figure 4A and Figure 4B, respectively. The Earth's AKR emission cone points tailward with partially overlapping northern and southern hemisphere cones (only the northern hemisphere cone at one frequency is shown in Figure 4A) and is not dependent on Earth's rotation or the location of the Moon.

One aspect of the Jovian DAM emission is that it is strongly coupled with the moon Io which has a thin atmosphere allowing an ionospheric current to connect field-lines from Jupiter creating a constant current and therefore a constant aurora and resulting CMI related radio emissions. These Io controlled DAM emissions produce hollow emission cones that move with Io around the planet which has an orbital period of about 42 hours. Io is in an elliptical orbit and is so close to the planet Jupiter that the energy from the very strong tidal forces is dissipated through volcanic activity on the moon that are so strong that a torus of escaping material is left in its wake that stretches around Jupiter. Alfven waves are set up in the Io torus that produce magnetospheric currents stretching all the way to the Jovian auroral regions that also trigger additional CMI emissions that produce a set of nested hollow emission cones. Near equatorial spacecraft, such as the Voyager 1 and 2 missions, observed the Io-DAM emissions as a series of arc-like structures in frequency-time spectrograms. The shape of the nested emission cones, in frequency-time spectrograms, are strongly controlled by the higher moments of the Jovian magnetic fields since the strongly right-hand polarized DAM radiation propagating from the source over these intense magnetic islands [Green, 1984] suffer significant refraction. In addition, Jupiter's moon Ganymede also produces aurora, not only in the upper atmosphere of Jupiter, but also in the very tenuous Ganymede atmosphere since that moon is the only one in the solar system that has been observed to currently generate its own magnetosphere [e.g., Wang et al., 2018]. The rather small Ganymede magnetosphere is antialigned with Jupiter which is similar to the lower left panel in Figure 1. These connected field lines also facilitate the exchange of atmospheric constituents.

In the case of both an exoplanet and exomoon with magnetospheres, we now have a new situation in which the exomoon would be controlling the location of a potential CMI emission cone and producing either a hollow or filled in emission pattern. From a distant radio observer, periodicities in an observed CMI emission cone pattern along with the radio emission frequency not only could point to the existence and strength of an exoplanet's magnetosphere but also the existence of an exomoon. The extent of the emission cone, ranging from completely hollow to completely filled-in provides additional information about the extent of the exoplanet's ionosphere. In this manner, the detection and analysis of CMI generated radio emissions may provide additional information as to the habitability of the exoplanet.

At this time, CMI radio emissions from exoplanet and/or exomoon magnetospheres have not had a confirmed detection, but there are three candidate systems under study [Turner *et al*., 2020]. In addition, there are a number of significant efforts underway to make these type of measurements [Griessmeier *et al.*, 2011]. Plans are underway to develop an extensive radio observatory on the far side of the moon, named FARSIDE [Burns *et al*., 2020]. This system is being designed to detect radio emission not only from solar system objects but from stellar radio bursts and the CMI emissions from potentially habitable exoplanets in the frequency range from 0.1-40 MHz, which encompasses both AKR and Io-DAM emission frequencies.

## 6. Conclusions
In this paper we modeled two dipole fields simulating the main field of the exoplanet and the exomoon when the exomoon was at several locations ranging from 4 to 18 $R_p$ from the exoplanet in a stellar wind environment. We take the Earth-Moon dipole strengths presented in Green *et al.* [2020] as our starting conditions illustrating a basic magnetic topology that would evolve over time.

It is well recognized that stellar winds from many types of stars can be extremely harsh to the point of blowing away atmospheres from rocky exoplanets residing within the HZs of the host stars. Therefore, coupled exoplanet-exomoon magnetospheres are likely to be of great importance for the study of exoplanets that are exposed to extreme stellar wind and particle radiation conditions. In order to understand the long-term evolution of exoplanetary atmospheres and their suitability for creating a habitable environment that may host life, we must understand not only the stellar environment, but also whether these planets and their associated moons have magnetic fields. Future detection of exoplanet-exomoon magnetic fields from the detection of CMI radio emissions will provide a wealth of new information that should draw our attention to these systems having a greater chance of habitability.

**Acknowledgments**
CD was supported by NASA grant 80NSSC18K0288.

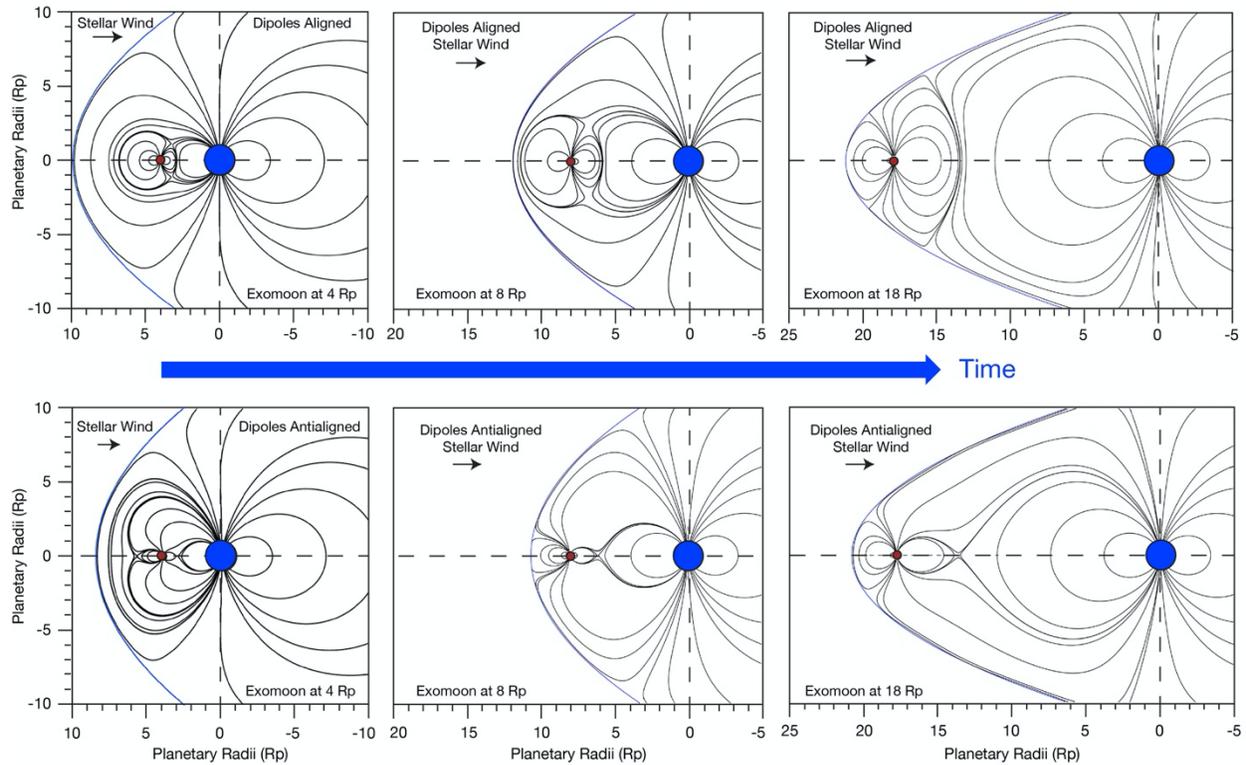

**Figure 1:** Simulation results for the exoplanet-exomoon coupled magnetospheres with the dipoles aligned (top panels) and dipoles antialigned (bottom panels) with three exomoon locations. This figure illustrates the evolution of the magnetic topology of an exoplanet-exomoon over a period of time delineated by the exomoon moving away from the exoplanet (expected to be a time period of 100's of millions of years).

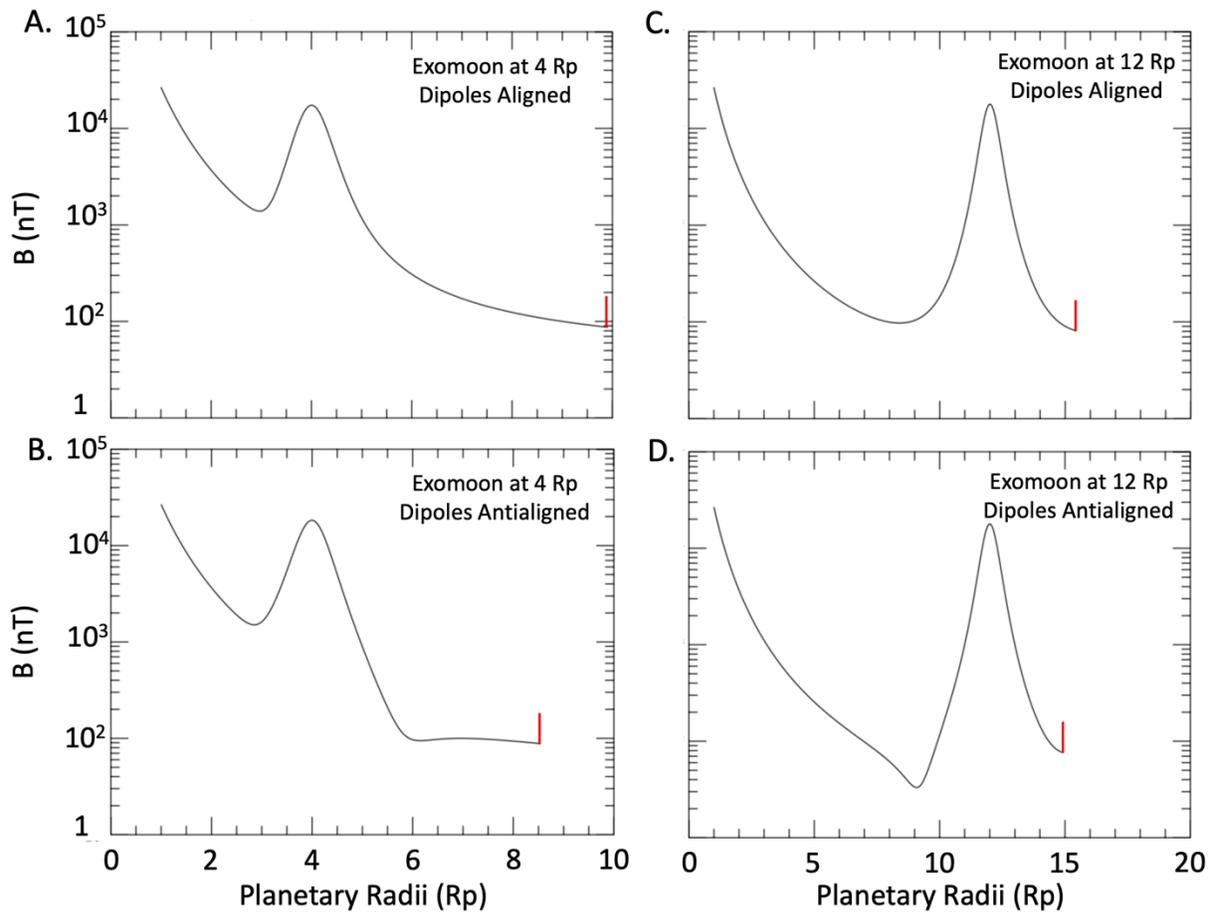

**Figure 2:** The intensity of the exoplanet-exomoon magnetic field from the exoplanet's surface through the magnetopause along the planet-star axis with the exomoon at 4 $R_p$ (panels A and B) and 12 $R_p$ (panels C and D).

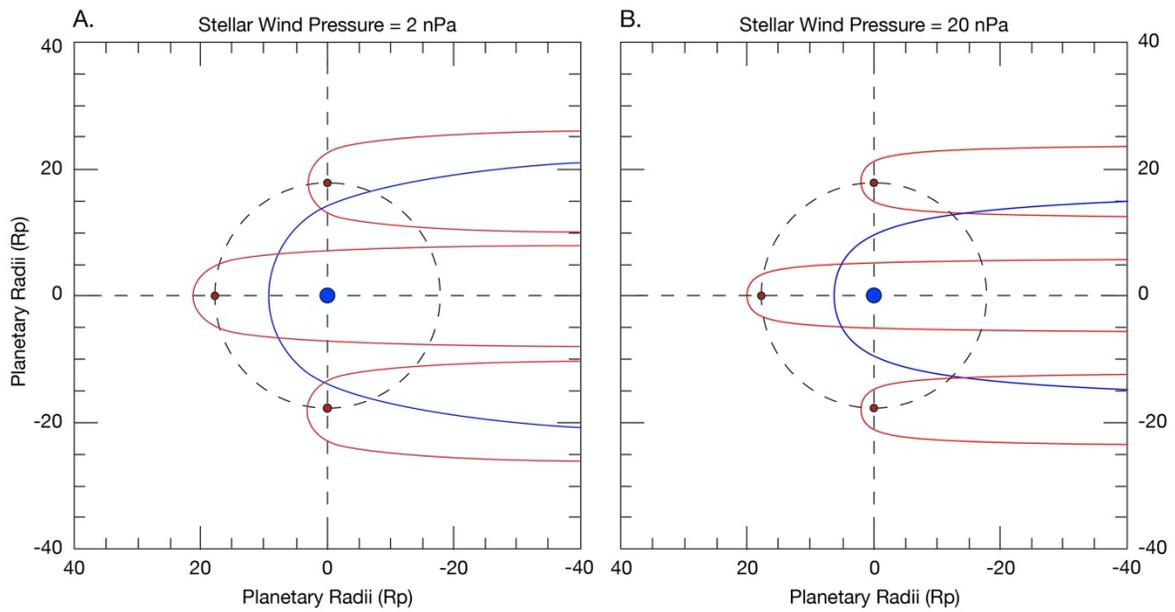

**Figure 3:** Polar view of magnetopause locations of an orbiting exomoon at 18 $R_p$ independent of the exoplanet magnetosphere under average (panel A) and extreme solar wind conditions (panel B).

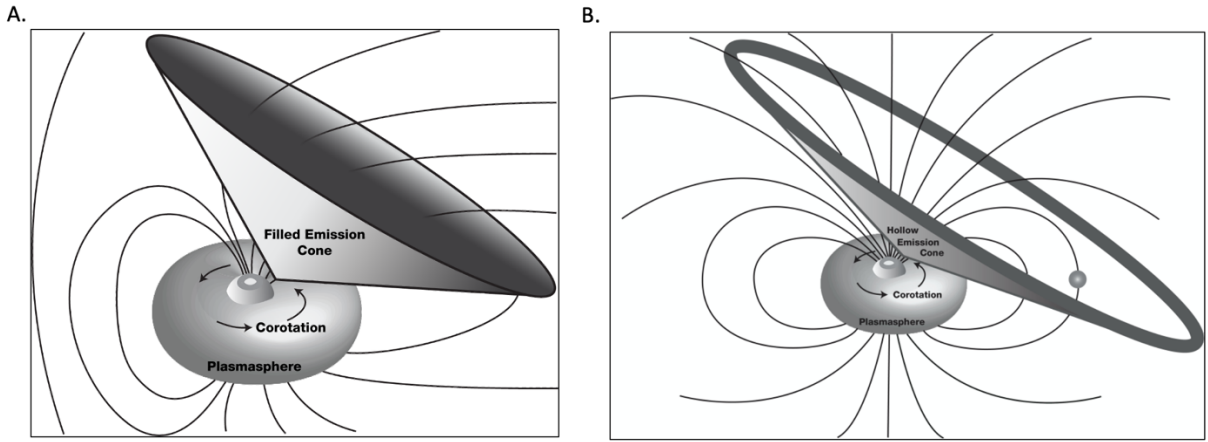

**Figure 4:** A schematic view of radio emission cones, at one emission frequency, modeled after AKR at Earth (panel A) and Io-control DAM at Jupiter (panel B).